\documentclass[journal=jacsat,manuscript=article]{achemso}
\setkeys{acs}{articletitle=true,keywords=true}
\usepackage[version=3]{mhchem} % Formula subscripts using \ce{}
\usepackage{graphicx} % Include figure files; subfigure
\usepackage{dcolumn} % Align table columns on decimal point
\usepackage{bm} % bold math
\usepackage{amsmath}
\usepackage{epsfig}
\usepackage{amssymb,amstext}
\usepackage{color}

\newcommand{\onlinecite}[1]{Ref.~[\hspace{-1 ex} \nocite{#1}\citenum{#1}]}
\title{Ultrafast dynamics of optically-induced heat gratings in metals}

\author{Yonatan Sivan}
\email{sivanyon@bgu.ac.il}
\affiliation{School of Electrical and Computer Engineering, Ben-Gurion University of the Negev, P.O. Box 653, Israel, 8410501}

\author{Marat Spector}
\affiliation{Department of Physics, Ben-Gurion University of the Negev, P.O. Box 653, Israel, 8410501}

\keywords{Ultrafast heat dynamics, Plasmonics, Transient gratings}

\begin{document}
\begin{abstract}
  Diffusion of heat in metals is a fundamental process which is crucial for a variety of applications of metal nanostructures. Surprisingly, however, {\em ultrafast} heat diffusion received only limited attention so far. Here, we show that heat diffusion can be made faster than $e-ph$ energy transfer rate, in which case, it dominates the spatio-temporal dynamics of the temperature. This enables the metals to overcome the conventional limitations of the nonlinear optical response of materials - it can be simultaneously fast and strong. As a specific example, we identify the underlying (femtosecond and few picosecond) time scales responsible for the generation and erasure of optically-induced transient Bragg gratings in thin metal films. Further, we show that heat diffusion gives rise to a significant nonlocal thermo-optic nonlinearity - it affects also the nonlinear optical response such that the overall change of the permittivity (hence, reflectivity of the transient grating) has a significant dependence also on the illumination period rather than only on the illumination intensity.
\end{abstract}

\section{Intro}

% no section numbers in ACS photonics, but there is section headings

Heat generation and dynamics is central to our understanding light-matter interactions in metals~\cite{Hohlfeld_Matthias_2000,Stoll_review,thermo-plasmonics-basics}, specifically for the purpose of separating thermal effects from (non-thermal) electronic effects~\cite{Dubi-Sivan,Dubi-Sivan-Faraday}. It is also critical for many applications~\cite{thermo-plasmonics-review,refractory_plasmonics} and specifically, for the correct interpretation of the role of heat in nanophotonic applications such as surface femto-chemistry~\cite{surface_fs_chemistry_Heintz,surface_fs_chemistry_Wolf} and plasmon-assisted photocatalysis~\cite{anti-Halas_comment,Ballistic_e_diffusion,Y2-eppur-si-riscalda,Liu_thermal_vs_nonthermal,Liu-Everitt-Nano-Letters-2019,Liu-Everitt-Nano-research-2019,Baffou-Quidant-Baldi}.

The standard model of heat generation and dynamics in metals is the well-known Two Temperature Model (TTM)~\cite{Two_temp_model} in which a distinction is made between the electron subsystem and the phonon subsystem (each assigned by a temperature, $T_e$ and $T_{ph}$, respectively); Further, the electron system is assumed to become thermalized instantaneously, hence, before significant energy transfer between the electrons and the phonons occurs. Under these conditions, the dynamic equations for the electron and phonon temperatures are coupled via a simple linear term, namely,
\begin{equation}\label{eq:TTM_simple}
\begin{split}
C_e(T_e) \frac{\partial T_e(t)}{\partial t}    &= - G_{e-ph}(T_e-T_{ph}) + p_{abs}(\vec{r},t), \\
C_{ph}   \frac{\partial T_{ph}(t)}{\partial t} &=   G_{e-ph}\left(T_e - T_{ph}\right).
\end{split}
\end{equation}
% Ad writes p_abs, but does not specify it.. delFatti 2000 does not
Here, $C_e$ and $C_{ph}$ represent the heat capacity of the electrons and phonons, respectively; $G_{e-ph}$ is the electron-phonon coupling factor, representing the rate of energy exchange between the electrons and the lattice; $p_{abs}$, is the density of absorbed photon power. These equations describe the decay of the electron temperature following an initial heating stage (which sometimes for convenience is even skipped altogether) due to energy transfer to the phonons; in conjunction, it describes the phonon heating. The TTM was originally introduced in~\onlinecite{Two_temp_model} and derived in detail from the semi-quantum mechanical Boltzmann equation for the few picosecond regime in~\onlinecite{delFatti_nonequilib_2000}. Unfortunately, as well known, the assumption underlying the TTM is never strictly valid - the thermalization of the electron subsystem extends roughly over the first picosecond after the excitation pulse exited the metal~\cite{non_eq_model_Ippen,non_eq_model_Lagendijk,Aeschliman_e_photoemission_review,delFatti_nonequilib_2000}.

As a remedy, using a unique model that ensures energy conservation, we derived in~\onlinecite{Dubi-Sivan,Dubi-Sivan-Faraday} an {\em extended} version of the TTM (referred to below as the eTTM) from the semi-quantum mechanical Boltzmann Equation. In this model, the early stages of the thermalization of the electron subsystem are accounted for via the exact electron distribution and the total energy of the non-thermal electrons; the latter is characterized by a fast (pulse-duration limited) rise time and slow decay time, corresponding to the thermalization of the electron subsystem%\footnote{In comparison, the source in the TTM has just the temporal profile of the absorbed power, $p_{abs}$. }
. This derivation confirmed the phenomenological models presented much earlier in~\onlinecite{non_eq_Fujimoto,non_eq_model_Ippen} as well as the classical derivations in~\onlinecite{non_eq_model_Carpene,nt_electrons}. In contrast to the TTM, the eTTM allows the electron subsystem to be non-thermal, and only assumes (e.g., by adopting the relaxation time approximation, or adopting a rigorous many-body formulation~\cite{delFatti_nonequilib_2000,Italians_hot_es,non_eq_model_Rethfeld_con}) that the electron subsystem can be described by some temperature - the one to which the electron system would have relaxed if it was isolated from the photons and phonons, see discussion in~\onlinecite{Dubi-Sivan,Dubi-Sivan-Faraday}. In that sense, within the validity conditions of the Boltzmann description, the eTTM is an exact coarse-grained description of the energy dynamics; it is valid at {\em all} times (in particular, also before all the non-thermal energy is depleted) and the temperature varies instantaneously upon absorption of photons as it reflects the total energy of the electron system.

% {\bf Te rises at the same rate, but drops on $\Gamma^{T_e}$ rate..}

% As a consequence, the thermal response of the irradiated system shows features that are not accounted for by the traditional TTM. In particular, the dependence the electronic temperature rise on the laser photon energy $\hbar\omega_{pump}$ and the identification of distinct heat sources for electrons and lattice are shown by~\eqref{eq:eqdynamicsNT}~-~\eqref{eq:TTM}.

% Renwen's model is fine in that sense!

% \footnote{Note that the differences between Eq.~(\ref{eq:eTTM_approx}) and the more conventional TTM extend only to the initial non-thermal regime, for which the temperature is anyhow not well defined. }.}

Most research on the heat dynamics in metals to date focussed on the derivation of the thermal properties, the details of the $e-ph$ coupling and the dynamics of the temperatures and electric fields. However, quite peculiarly, early studies regularly did not account for heat diffusion. This approach suited most of the configurations studied experimentally~\cite{Brorson1987,non_eq_model_Lagendijk}. Indeed, on one hand, nanometric metal particles or thin films are characterized by a uniform electric field and hence, uniform temperatures. In fact, the temperature is uniform even for metal nanostructures extending to several tens or even hundreds of nanometers, due to the strong heat diffusion in metals~\cite{thermo-plasmonics-basics,Un-Sivan-size-thermal-effect}. Furthermore, whenever a large beam was used for illumination, diffusion was a slow process (see Eq.~(\ref{eq:taudiff}) below) occurring only on the edges of the illumination spot, away from where the processes of interest occurred~\cite{non_eq_model_Ippen,non_eq_model_Lagendijk,delFatti_nonequilib_2000}. However, on the other hand, it is easy to appreciate that there are several scenarios in which heat diffusion cannot be neglected. These include, for example, thick metal layers into which light penetration is far from being complete~\cite{Brorson1987,Hohlfeld_Matthias_2000,Rotenberg_PRB_09,Planken_buried_gratings_2018}, metal-dielectric composites~\cite{Khurgin-diffusive-switching,Peruch-adv-opt-materials-2017}, diffusion of a localized heat spot in a thin film (studied in~\onlinecite{ICFO_Sivan_metal_diffusion}), highly non-uniform illumination of micron-scale metal objects etc.. In such cases, the heat diffusion is expected to cause the strongly illuminated regions to reach lower maximal temperatures (compared to the diffusion-free case) and the weakly-illuminated regimes to initially get hotter (before cooling down with the rest of the system due to heat transfer to the environment), see e.g.~\onlinecite{Peruch-adv-opt-materials-2017}; heat diffusion may also affect the overall time scales for the dynamics in a non-trivial way (e.g., to increase the electron temperature rise time~\cite{Lalanne_Narang_Nat_Comm_2017}, to tailor the slower $e-ph$ energy transfer rate~\cite{Nicholls-Zayats-2019}, or more generally, to affect the overall dynamics due to the relevant modal response~\cite{Peruch-adv-opt-materials-2017}).

\begin{figure}[H]
  \centering{\includegraphics[width=4cm,width=8cm]{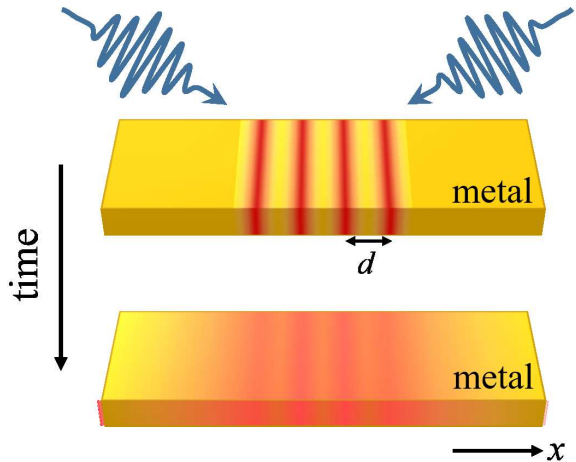}}% {schematics.png}} - replaced just for the ArXiv submission
  \caption{(Color online) A schematic illustration of an optically-induced temperature grating, and the dynamics of its contrast ($\Delta T(t)$, the temperature difference between red and yellow regions). }
  \label{fig:schematics}
\end{figure}

In this manuscript, we focus on the specific case of the diffusion of a periodic heat pattern (aka transient Bragg gratings, TBGs) in a thin metal film. In this scenario (depicted schematically in Fig.~\ref{fig:schematics}), two pump pulses are interfered onto a metal film to create a periodic pattern of absorption; in turn, the generated heat and the temperature dependence of the metal permittivity gives rise to a periodic permittivity modulation that outlives the pulse, but self-erases due to heat diffusion. Such a pattern can enable transient reflectivity of guided modes in the metal film that can be turned on and off on a sub-picosecond time scale, as for free-carrier gratings~\cite{Sivan-COPS-switching-TBG}. This property is appealing for ultrafast switching applications, as it may enable switching speeds significantly faster than what is commonly achieved with free-carrier generation in semiconductors~\cite{Lipson_review}. To the best of our knowledge, TBGs in metal films were studied before only in the context of the slow (nanosecond and microsecond) phonon dynamics (see, e.g.,~\onlinecite{Matthias-metal-TBGs-1994,Matthias-metal-TBGs-1995,Shen-metal-TBGs-1998,Wada-metal-TBGs}), where the electron thermalization and sub-picosecond temperature dynamics can be conveniently ignored. Studies of faster dynamics were performed in~\onlinecite{Maznev-Nelson-2011} for time scales of several hundreds of picoseconds and in~\onlinecite{Planken_buried_gratings_2018} on a scale of few tens of picoseconds. % ($6 \mu$m gratings in layered metal structures). % 30fs pulses.. modelling without thermalization, only vertical diffusion. no time scale analysis. % {\bf long grating papers -- which?? -- also study the acoustic behaviour..}
The neglect of the subpicosecond dynamics might have also originated from the un-availability of a proper theoretical tool valid in that temporal range. An exception is the study of Ivanov {\em et al.}~\cite{Ivanov_Phys_Rev_Appl_2015} which was, however, focussed on employing the periodic illumination pattern for patterning the surface of a metal film.

Accordingly, using the (all-time valid) eTTM, we focus in the current manuscript on the %account for heat diffusion and the demonstration of its control via the illumination inhomogeneity (namely, the spatial extent of the pump, the penetration depth, the period of the pump pattern etc.).
subpicosecond and few picosecond spatio-temporal heat dynamics in optically-induced TBGs in metals, specifically, on the interplay between the numerous time scales - the pump duration (via $p_{abs}$), the thermalization rate (defined as $\Gamma^{NT}$, see Eq.~(\ref{eq:Gamma_NT}) below), $e-ph$ energy exchange rate ($\Gamma^{T_e} = G/C_e$, see Eq.~(\ref{eq:TTM_simple})) on one hand, and the diffusion time $\tau_{diff}$ on the other hand. Based on standard (i.e., single temperature) heat equations, the latter is defined using the second-order (spatial) moment of the solution of the diffusion equation subject to the periodic modulation, namely,
\begin{equation}\label{eq:taudiff}
\tau_{diff} \equiv \frac{d^2}{4\pi^2 D_e}, \quad\ D_e \equiv \frac{K_e}{C_e}.
\end{equation}
Here, $D_e$ and $K_e$ are the diffusion coefficient and thermal conductivity of the metal electrons; $d$ is a characteristic length scale associated with the initial temperature non-uniformity; here, it is simply the illumination period. The magnitude of the diffusion constant $D_e$ of metal electrons is approximately $D_e\approx 100 - 300\ cm^2 s^{-1}$ (as e.g., demonstrated experimentally~\cite{ICFO_Sivan_metal_diffusion}). % ,  in the current electron temperature regime.%\footnote{The heat capacity of the metal is defined as the sum of the heat capacity of the lattice, $C_{ph}$, and the heat capacity of the electrons, $C_e$, where $C_{ph} / C_e \approx 100$~\cite{Ashcroft-Mermin}. This may serve as a possible explanation why some articles~\cite{Diff,Luther_Davies} show values for the diffusion coefficient that are 2 orders of magnitude lower compared with those shown in Fig.~\ref{fig:diff_time}. {\bf don't these papers say explicitly that D is of the phonons? }}.
With sufficiently strong inhomogeneities of the illumination (specifically, small periods, in the current context), the heat diffusion can become as fast as a few hundreds of femtoseconds.

We show that due to the non-trivial interplay between the temporally- and spatially-{\em nonlocal} nature of the thermo-optic response of metals (namely, thermalization and $e-ph$ energy transfer vs. heat diffusion), heat diffusion dominates the dynamics when it is faster than $e-ph$ heat transfer ($\sim 1.25 ps$ for Au). This physical scenario explains why the standard trade-off of the nonlinear optical response of materials between speed and strength~\cite{Boyd-book} can be surpassed by metals (see discussion in~\onlinecite{Zheludev-metal-nlty}), enabling the thermo-optic nonlinearity in metals to be both fast and strong. When the diffusion time is slower, the dynamics is dominated by the {\em spatially-local but temporally-nonlocal} $e-ph$ energy transfer. In this regime, the gratings do not get totally erased, but rather persist for many picoseconds. This means that, contrary to our initial expectations, TBGs in metals are not ideal as self-erasing TBGs in ultrafast switching applications, except maybe in the extreme spatially-nonlocal limit.

Finally, we also show that the spatial nonlocal nature of the nonlinear thermo-optic response of the bulk metal (not to be confused with the {\em structural} nonlocality in metal-dielectric composites, see~\onlinecite{Zayats_Nonlocal_nlty}) affects the overall magnitude of the nonlinear response. This newly observed effect might have been playing a role in the multiplicity of reported values of the intensity-dependent optical nonlinearity of metals and their composites (see e.g.,~\onlinecite{Boyd-metal-nlty,NL_plasmonics_review_Zayats_2012}) and provides yet another motivation to avoid describing the nonlinear optical response by a nonlinear susceptibility, a concept which was derived in the context of spatially- and temporally-local (i.e., Kerr) electronic nonlinearities.

While the results of the current manuscript rely on a rather generic ``minimal'' model applied to metals only, in its last part we describe several aspects of the model that can be improved for more accurate modelling of specific metal systems, and refer briefly to the possibility of observing similar effects in other materials systems like semiconductors, 2D materials etc..

% This case embodies a qualitative change in the optical response due to diffusion and is promising for ultrafast switching - this we might want to remove...

% {\bf \cite{Lalanne_Narang_Nat_Comm_2017} - longer effective $e-ph$ relaxation because of heat flow away from the tip. they have a term for non-thermal e's to phonons..., % - taper something}
% surface ablation, see~\cite{Yoann_Levy_2016} and refs therein - there are earlier studies of this..
% }

% heat diffusion is extremely fast in metals, with sub-picosecond diffusion times, yet, it turns out to have only a relatively limited effect on the overall dynamics.

% nlty of a focussed spot and various 4 wave mixing processes with different wavelengths (hence, periods) will be different - but the latter is coherent...

% {\bf could it be that the local time scales were effective in~\cite{ICFO_Sivan_metal_diffusion} but we did not refer to them?  }

% This paper is organized as follows. In Section~\ref{sec:eTTM} we introduce and solve the eTTM, and discuss the numerical results. We show that ... . In Section~\ref{sec:epsilon}, we compte the spatio-temporal dynamics of the permittivity induced by the temperate dynamics. In Section summary we summarize...

\section{Spatio-temporal dynamics of the electronic and phonon temperatures}\label{sec:eTTM}
\subsection{Model - the extended Two Temperature Model (eTTM)}

We adopt here the extended Two Temperature Model (eTTM)~\cite{Dubi-Sivan,Dubi-Sivan-Faraday} which describes the spatio-temporal dynamics of $\mathcal{U}_e^{NT}$, the non-thermal (NT) electron energy, as well as of the electron and phonon (lattice) temperatures. In this approach, $\mathcal{U}_e^{NT}$ is determined by~\cite{non_eq_model_Ippen,non_eq_model_Carpene,Dubi-Sivan,Dubi-Sivan-Faraday}$^;$\footnote{Note that the eTTM does not capture the increase of rate of energy transfer from electrons to the lattice during the thermalization time, as discussed in~\onlinecite{non_eq_model_Lagendijk}; however, this effect should have, at most, a modest quantitative effect on the issues discussed in the current work.}
\begin{equation}\label{eq:dynamicsNT}
\frac{\partial \mathcal{U}_e^{NT}(\vec{r},t)}{\partial t} = - (\Gamma_e + \Gamma_{ph}) \mathcal{U}_e^{NT}(\vec{r},t) + \langle p_{abs}\left(\vec{r},t\right) \rangle. % last term was denoted by \alpha I in Marat's earlier work
\end{equation}
% where the first term on the RHS describes the macroscopic $e-e$ and $e-ph$ scattering rates, represented by $\Gamma_e,\Gamma_{ph}$, respectively%\footnote{Note that these rates differ from the microscopic ones denoted by $\gamma_{e-e}$ and $\gamma_{e-ph}$~(\ref{eq:tautot}) that represent (effective) single electron collision events. }.
Here, $\Gamma_e$ represents the rate of exchange of energy between the non-thermal electrons and the thermal ones, or in other words, it represents the thermalization rate of the electron distribution as a whole~\cite{Aeschliman_e_photoemission_review}. It was computed and measured to be on the order of several hundreds of femtoseconds (see e.g.,~\onlinecite{non_eq_Fujimoto,non_eq_model_Lagendijk,nt_electrons,GdA_hot_es} to name just a few of the studies of the thermalization). In~\onlinecite{non_eq_model_Ippen} it was found experimentally that $\Gamma_e$ is independent of the laser fluence in the range of $2.5-200 \frac{\mu J}{cm^2}$, corresponding to an electron temperature rise of up to about $200$K. %, relevant to all host media because on the picosecond time scales, the heat transfer to the environment is negligible.. %\equiv \Gamma_e
Similarly, $\Gamma_{ph}$ represents the rate of energy transfer between the non-thermal electrons and the phonons. Due the similarity in the physical origin, it is set to the same value chosen for the energy transfer rate between the thermal electrons and the phonons ($\Gamma^{T_{ph}}$), see below. Together they form the total decay rate of the NT energy (also referred to as the thermalization rate) denoted by
\begin{equation}
\Gamma^{NT} = \Gamma_e + \Gamma_{ph}. \label{eq:Gamma_NT}
\end{equation}
The second term on the RHS, $\langle p_{abs} \rangle$, describes the time-averaged power density of absorbed photons. Using the Poynting theorem for dispersive media and assuming that all the absorbed photon energy is converted into heat, % (see e.g.,~\cite{Stoll_review,Langbein_PRB_2012})
together with the assumption of a slowly varying envelope approximation in time for the electric field, it can be written as%\footnote{For further details see Appendix~\ref{sec:dissipation}.}
% by IW - time averaging gives 1/2..; E is time dependent - Marat/Landau;
\begin{equation}\label{eq:p_abs}
\langle p_{abs}(\vec{r},t) \rangle = \frac{1}{2} \epsilon_0 \epsilon''_m \omega_{pump} \langle |\vec{E}\left(\vec{r},t\right)|^2 \rangle,
\end{equation}
where $\epsilon_m$ is the metal permittivity (and $\epsilon_m''$ is its imaginary part), $\omega_{pump}$ is the pump frequency, $\vec{E}$ is the local electric field and $\langle \rangle$ stands for time-averaging over the period; note that $\langle p_{abs}(\vec{r},t) \rangle$ maintains the time dependence of its envelope. % IWU - they used the correct Ce and Cph. % in fourier domain - convolve the E with itself..% Marat's derivation wrong - should have involved $\omega_0$...
% Since $\langle p_{abs}\rangle$ has the same time-dependence as the pulse envelope, the generation rate of the NT energy (as well as the rise time of the electron temperature, see below) is governed by the optical pump pulse duration $\tau_{pump}$.

The equations for the temperatures in the original TTM are
\begin{equation}\label{eq:TTM}
\begin{split}
C_e(T_e) \frac{\partial T_e(\vec{r},t)}{\partial t} &= - G_{e-ph}(T_e-T_{ph}) + \Gamma_e \mathcal{U}_e^{NT}(\vec{r},t), \\
C_{ph}   \frac{\partial T_{ph}(\vec{r},t)}{\partial t} &= G_{e-ph}\left(T_e - T_{ph}\right) + \Gamma_{ph} \mathcal{U}_e^{NT}(\vec{r},t).
\end{split}
\end{equation}
In these coupled heat equations, $\mathcal{U}_e^{NT}$ serves as the heat source, and $C_e$ and $C_{ph}$ are the heat capacities of the electrons and the lattice; $G_{e-ph}$ (mentioned above) represents the rate of energy exchange between the electrons and the lattice which occurs on time scales defined as~\cite{non_eq_model_Lagendijk} $\Gamma^{T_e} \equiv G_{e-ph} / C_e$ and $\Gamma^{T_{ph}} \equiv G_{e-ph} / C_{ph}$, respectively; Since $\Gamma^{T_e} \gg \Gamma^{T_{ph}}$, the latter is essentially negligible. These equations neglect heat coupling to the environment and consequent cooling, as it occurs typically on time scales much longer than considered in the current work.

In order to account for heat diffusion, we add to the standard equations above a standard heat diffusion term. We emphasize that as there is currently no self-consistent derivation of such terms from a first-principles model like the Boltzmann equation or a Density Matrix Formulation, the insertion of heat diffusion should be considered as being {\em purely phenomenological}. Following the recent experimental measurement of heat diffusion of a single heat spot in a thin metal film~\cite{ICFO_Sivan_metal_diffusion}, we assume that the diffusion coefficient of the non-thermal and thermal electrons is the same. This assumption is in line with the findings of Hohlfeld {\em et al.}~\cite{Hohlfeld_Matthias_2000} for vertical diffusion in thick metal films. Thus, the model equations above are modified to
\begin{equation}\label{eq:TTM}
\begin{split}
\frac{\partial \mathcal{U}_e^{NT}(\vec{r},t)}{\partial t} &= \frac{1}{C_e(T_e)} \nabla[K_e(T_e,T_{ph})\nabla \mathcal{U}_e^{NT}] - (\Gamma_e + \Gamma_{ph}) \mathcal{U}_e^{NT}(\vec{r},t) + \langle p_{abs}\left(\vec{r},t\right) \rangle. \\
C_e(T_e) \frac{\partial T_e(\vec{r},t)}{\partial t} &= \nabla[K_e(T_e,T_{ph})\nabla T_e] - G_{e-ph}(T_e-T_{ph}) + \Gamma_e \mathcal{U}_e^{NT}(\vec{r},t), \\
C_{ph}   \frac{\partial T_{ph}(\vec{r},t)}{\partial t} &= \nabla[K_{ph}(T_{ph})\nabla T_{ph}] + G_{e-ph}\left(T_e - T_{ph}\right) + \Gamma_{ph} \mathcal{U}_e^{NT}(\vec{r},t).
\end{split}
\end{equation}
Here, $K_e$ and $K_{ph}$ are the thermal conductivities of the electrons and the lattice. Note, however, that due to the relative smallness of the phonon diffusion $K_{ph}$, it is henceforth neglected.

The model described above is used below to characterize the ultrafast heat diffusion in the periodically-illuminated thin metal film. We emphasize that in order to maintain simplicity and without loss of generality, inclusion of more advanced modelling (e.g., accounting more accurately for thermalization~\cite{Aeschliman_e_photoemission_review,Wilson-Coh}, for the temperature dependence of the optical and thermal parameters~\cite{Sivan-Chu-high-T-nl-plasmonics,Gurwich-Sivan-CW-nlty-metal_NP} (see also discussion below), for potential anisotropy of the parameters, for differences between different metals, for quantum mechanical effects etc.) is intentionally avoided in order to keep the generality and simplicity of the discussion. Those effects are discussed briefly at the end of manuscript and would be included in future studies.

\subsection{Numerical results}

% I do not understand Marat's boundary condition. it reads K dT_e/de = Te - Teq; the units do not make sense. with numbers this is dT_e/dx ~ 100K / (315 W / m K)

We now solve Eqs.~\eqref{eq:TTM} for the case in which the metal film is optically thin (compared to the optical penetration (skin) depth into the metal) and the illumination is periodic. Accordingly, we assume that all quantities depend only on $x$ and $t$ and look at the dynamics within a single period $d$. In addition, we set the pump pulse to be the shortest time-scale in the system (i.e., $\tau_{pump} \ll \tau_{diff}, 1/\Gamma^{NT}, 1/\Gamma^{T_e}$, ...); the latter choice is motivated a-posteriori by the weak dependence on the pump duration observed in further simulations (not shown). The parameters used in the solution of the {\em eTTM}~\eqref{eq:TTM} are given in Table~\ref{tab:parTTM}, suitable for Au; the parameters for Ag and Al are quite similar. We assume that the pump illuminates the sample near its plasmon resonance with a local field of $\sim 30 MV/m$.\footnote{Note that we avoid specifying the local intensity, as it is a somewhat improper quantity to use in the context of metals. Indeed, the negative real part of the permittivity causes the fields within the metal to be primarily evanescent, hence, not to carry energy (such that the Poynting vector, hence, intensity vanish, at least in the absence of absorption). Instead, we use the local density of electromagnetic energy, by specifying the local {\em electric} field, which is easy to connect to the incoming field. } Importantly, the comparisons below are performed for the same local field (hence, absorbed power density) within a single unit-cell\footnote{This allows us to avoid the complication associated with the strong angle and frequency dependence of the relation between the incoming and local fields. }.

Solving the complete eTTM requires knowledge of the temperature dependence of all parameters. % While this dependence is fairly simple for the thermal properties, the temperature dependence of the permittivity (and the electric field) is far more complicated.
In the current manuscript, we neglect the temperature-dependence of all parameters (most notable of which is of the heat capacity), as appropriate for low intensities whereby the temperature rise is modest~\cite{Gurwich-Sivan-CW-nlty-metal_NP}. Due to the brief duration during which the system temperatures are more than several hundreds of degrees above room temperature, any changes associated with such temperature dependence make only modest quantitative changes to the results shown below.
% we neglect the temperature dependence of $\Gamma_e$ (whose value CHANGE IF I DECIDE TO KEEP GAMMA-PH is anyhow phenomenological)%{\bf can we say anything about their T dependence? we can say that the microscopic collision rate, near the Fermi energy scales quadratically with $T_e$.}
%, $\langle p_{abs} \rangle$ (i.e., neglecting the thermo-optic nonlinearity; for CW illumination, this was shown in~\cite{Gurwich-Sivan-CW-nlty-metal_NP} to be valid for temperature rises lower than $100K$), $C_e$ {\bf (see Eq.~(\ref{eq:heatcap})?? above)}, $C_{ph}$\footnote{In general, the lattice heat capacity $C_{ph}$ depends on the lattice temperature $T_{ph}$ through the Debye model~\cite{Ashcroft-Mermin}. However, in the current temperature range, $T_{ph} > T_D$, we can neglect its temperature dependence.}, $K_e$ {\bf (see Eq.~(\ref{eq:thermcon})?? above)} %, $K_l$
%and $G_{e-ph}$ (shown to be temperature-independent till $3000$K~\cite{Brown_PRB_2016}).
% The corresponding room temperature quantities are denoted by the subscript 0.

Under these conditions%\footnote{Valid even including temperature-dependence of $\langle p_{abs} \rangle$, but not in the decay coefficients. }
, the solution for the non-thermal energy dynamics was shown in~\onlinecite{Langbein_PRB_2012} to be
\begin{equation}\label{eq:U_NT_solution}
\mathcal{U}_e^{NT}(x,t) = \frac{\sqrt{\pi} \tau_{pump}}{2} e^{{\Gamma^{NT}}^2 \tau_{pump}^2/4 - \Gamma^{NT} t} \left[1 + erf\left(\frac{t}{\tau_{pump}} - \frac{\Gamma^{NT} \tau_{pump}}{2}\right)\right] p_{abs,0},
\end{equation}
where (following~\onlinecite{Sivan-COPS-switching-TBG})
\begin{equation}
p_{abs,0} = \left[1 + \cos (2 \pi x / d) e^{- \frac{t}{\tau_{diff}}}\right] \frac{\epsilon_0 \epsilon''_m \omega_{pump}}{2} max_{x \& t} |\vec{E}\left(x,t\right)|^2.
\end{equation}
This shows that the grating contrast decays exponentially at the diffusion time scale without modifying the average spatial absorption profile.
% {\bf add details in appendix?}

For a short pump pulse, $\mathcal{U}_e^{NT} \sim \delta(t) e^{- \Gamma^{NT} t}$, such that the pump pulse and NT energy peak together. For a longer pump pulse, the NT energy becomes maximal later than the peak of the pump pulse. % this is due to thermalization - if there was none, the peak of the NT energy would be at infinity.
Furthermore, the non-instantaneous thermalization gives rise to a temporally smeared source for the electron temperature % works on the laptop...
compared to the source appearing in the TTM~\cite{Two_temp_model} (which is just $\langle p_{abs}\rangle$~(\ref{eq:p_abs})).

% We study different values of electron heat diffusion time $\tau_{diff}$.

% see discussion with Marat regarding the difficulties with the sech profile.
% {\bf Marat's solution decays on the tau-pump scale.. the numerics decays on $\Gamma^{NT}$ scale.. }

% Assuming small variation of the electron and lattice temperatures with respect to their equilibrium temperature, we can linearize Eqs.~\eqref{eq:TTM}, namely,

% {\bf Note that the time scales responsible for the decay of the grating contrast namely $\Gamma^{T_e}$, $\Gamma^{T_{ph}}$, $\tau^{-1}_{diff}$ appear together in Eq.~\eqref{eq:electrontempgrat}. % In contrast, the time scales responsible for the rise of the grating contrast namely $\tau_{pump}$, and $\Gamma^{NT}$ appear separately in Eq.~\eqref{eq:electrontempgrat}. }

%{\bf Marat - the following dealt only with the TTm, right? if so, remove.. Another attempt to solve the {\em eTTM} analytically, for the case of uniform excitation, has been presented in~\cite{non_eq_model_Lagendijk,DelFatti3}. Although the authors take into account the temperature dependence of the electronic heat capacity $C_e\left(T_e\right)$, they ignore the {\em non} equilibrium stage. }

% {\bf magnitude of $\mathcal{U}_e$ not much greater than $\mathcal{U}_e^{NT}$!}

\begin{table}[H]
  \begin{center}
  \begin{tabular}{ | l | l | l |  l |}
  \hline
  Parameter  & Value & Units & Reference \\ \hline
  $C_e$  & $2 \cdot 10^4$ & $Jm^{-3}K^{-1}$ &~\onlinecite{Ashcroft-Mermin}\\ \hline % Marat wrote 2.1, which is incorrect ratio even according to his numbers.
  $C_{ph}$   & $2.5 \cdot 10^6$ & $Jm^{-3}K^{-1}$ &~\onlinecite{Ashcroft-Mermin}\\ \hline
  $G_{e-ph}$        & $2.5 \cdot 10^{16}$    & $Jm^{-3}K^{-1}s^{-1}$ &~\onlinecite{el_ph_constant}\\ \hline
  $\Gamma_e$ & $2 \cdot 10^{12}$& $s^{-1}$ &~\onlinecite{non_eq_model_Ippen}\\ \hline
%   $\Gamma_{ph}$     & $0$& $s^{-1}$ &~\cite{Dubi-Sivan-Faraday} \\ \hline
  $\Gamma^{NT}$     & $2 \cdot 10^{12}$& $s^{-1}$ &~\onlinecite{non_eq_model_Ippen}\\ \hline
  $\Gamma^{T_e}$    & $1.25 \cdot 10^{12}$& $s^{-1}$ &~\onlinecite{non_eq_model_Ippen}\\ \hline
  $\Gamma^{T_{ph}}$ & $10^{10}$& $s^{-1}$ &~\onlinecite{non_eq_model_Ippen}\\ \hline
  $K_e$  & $315$& $W m^{-1} K^{-1}$  & ~\onlinecite{Ashcroft-Mermin} \\ \hline
  $\tau_{pump}$  & $100$& $fs$&  -----\\ \hline
  $\lambda_{pump}$  & $500$& $nm$&  -----\\ \hline
%   $I_{pump}$  & $0.45$ & $GW/cm^2$ &  -----\\ \hline
%   $n_d$      & $1$  & ------  & ----- \\ \hline
  \end{tabular}
  \caption{Parameters used in the numerical solution of the {\em eTTM}~(\ref{eq:dynamicsNT})-\eqref{eq:TTM}.}\label{tab:parTTM}
\end{center}
\end{table}

\begin{figure}[H]
\centering{\includegraphics[width=18cm]{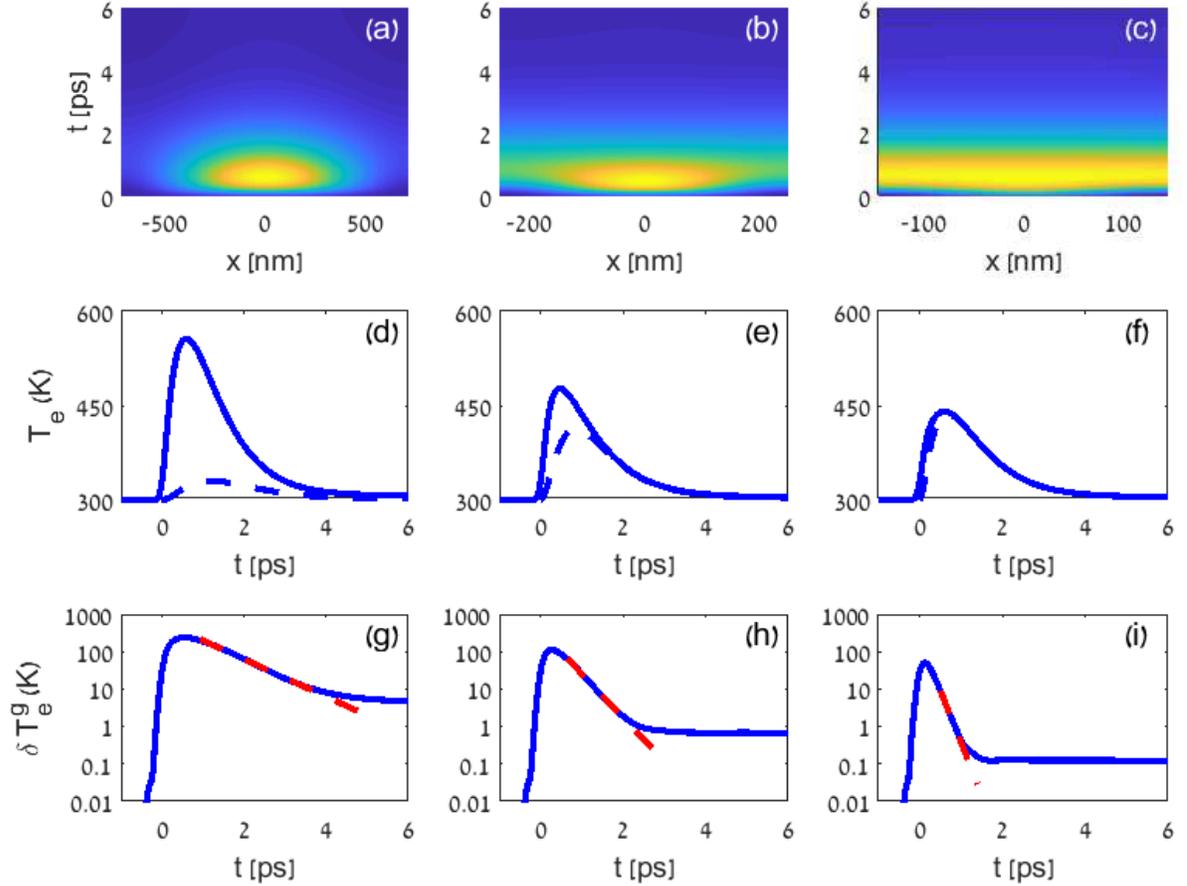}}
  \caption{(Color online) Spatio-temporal dynamics of the temperature of the electrons $T_e$ for an optically thin metal film when the diffusion time (Eq.~\eqref{eq:taudiff}) and period $d$ are (a) $\tau_{diff} = 3.6$ps and $d = 1.45\mu m$, (b) $\tau_{diff} = 430$fs and $d \approx 0.5\mu m$ and (c) $\tau_{diff} = 150$fs and $d \approx 0.3\mu m$; temperature range plotted is $300$K to $600$K in all plots. (d)-(f) Cross-sections of the maps directly above, specifically, $T_e(x_{max})$ (solid line), and $T_e(x_{min})$ (dashed line). (g)-(i) The corresponding grating contrasts in log scale (blue solid lines), along with the exponential fits of the decay stage (dashed red lines). } \label{fig:dynamicsTel}
\end{figure}

Figs.~\ref{fig:dynamicsTel}(a)-(c) show maps of the spatio-temporal dynamics of $T_e$ for different illumination periods $d$ (hence, different diffusion times). % ; all cases involve the same total power incident over the unit-cell.
Figs.~\ref{fig:dynamicsTel}(d)-(f) show the corresponding cross-sections of $T_e$ at $x_{max} = 0$ (where the illumination intensity is maximal), and at $x_{min} = d/2$ (no illumination). In all cases, $T_e(x_{max})$ initially builds up due to thermalization (i.e., due to energy transfer from the NT electrons to thermal ones); This occurs on a time scale $\Gamma_e^{-1}$ which is longer than the pump duration, hence, this is referred to as a temporally-nonlocal effect~\cite{Biancalana_NJP_2012}. The maximal temperature rise with respect to room temperature is $\sim 85\%$, $\sim 60\%$, $\sim 45\%$ for the three cases we simulated. The temperature at $x_{max}$ then gradually decreases due to both $e-ph$ energy transfer and electron heat diffusion. The latter effect causes $T_e(x_{min})$ to build up as well, despite not being initially illuminated (hence, this is a spatially-nonlocal effect). Figs.~\ref{fig:dynamicsTel}(g)-(i) show the electron temperature grating contrast, defined as $\delta T_e^g \equiv T_e(x_{max}) - T_e(x_{min})$. % for $\Gamma^{T_e} > \Gamma^{NT}$, the electron T does not grow as much
We can identify 2 limits~-
\begin{itemize}
  \item The {\bf spatially-local limit}, $1/\Gamma^{NT}, 1/\Gamma^{T_e} %\approx 4.5
  \ll \tau_{diff}$, i.e., when the electron heat diffusion is much slower than the local relaxation of $T_e$ (see Figs.~\ref{fig:dynamicsTel}(a)~\&~(d)). In this case, most of the electron heat is transferred to the lattice before any significant amount of heat reaches the minimum point from the maximum point. As a result, the high temperature region is localized and the electron temperature grating contrast is maximal. % $3.6/(1/1.2)$

  \item The {\bf spatially-nonlocal limit}, $1/\Gamma^{NT} %\approx 0.18
  , 1/\Gamma^{T_e} \gtrsim \tau_{diff}$, i.e., when the electron heat diffusion is faster than the local energy relaxation of NT electrons (see Figs.~\ref{fig:dynamicsTel}(c)~\&~(f)). In this case, the fast diffusion of electron heat limits the heating of the illuminated regions, while promoting the heating of the un-illuminated regions (as also observed previously in~\onlinecite{Peruch-adv-opt-materials-2017}). % almost uniform temperature of electrons $T_e$ in space even before the electron energy is transferred to the lattice.
   Consequently, the high temperature region is relatively delocalized and the electron temperature grating contrast, $\delta T_e^g$, is minimal. % $0.15/(1/1.2)$

  % Consequently, the values of $T_e$ are smeared down and as a result the maximum values of $T_e(x_{max})$ and of $T_e(x_{min})$  are approximately the same for different values of $\tau_{diff}$  as shown in Figs~\ref{fig:dynamicsTe150fs}~\&~\ref{fig:dynamicsTe150fsmaxmin}).

\end{itemize}

For any choice of parameters between these two extremes (e.g., Figs.~\ref{fig:dynamicsTel}(b)~\&~(e)), a non-negligible amount of heat is transferred from the maximum point to the lattice before it reaches the minimum point. % Thus, the electron temperature grating contrast takes intermediate values in this case (see Figs.~\ref{fig:dynamicsTe600fs}~\&~\ref{fig:dynamicsTe600fsmaxmin}) as compared to the case above. %that is shown in Figs~\ref{fig:dynamicsTe4ps}~\&~\ref{fig:dynamicsTe4psmaxmin}.

% {\bf Maximal contrast builds up in the early stages of the dynamics, during which the unilluminated regions did not yet heated up. at later stages, the contrast decays... maybe this is too detailed...}

Further simulations show that as the pump duration is increased (at a fixed pulse energy), the temporal maximum of $T_e(x_{max})$ drops only slightly and the temperature at the minimum remains roughly the same, so that the electron temperature grating contrast decreases only slightly as well. Accordingly, we do not pursue the study of longer pulses.

% {\bf peak of the grating contrast equals the peak of Te for large $d$, otherwise at slightly earlier times.}

% {\bf need this? maybe after the numerical results in the context of a discussion of the analytic solution? The non-monotonic {\bf nonlinear?} dependence of $T_e$ on $\tau_{diff}$ is attributed to the nonlinear dependence of $T_e$ on $\tau_{diff}$\footnote{For further details see Appendix~\ref{sec:solution}}. }

\begin{figure}[H]
  \centering{\includegraphics[width=16cm]{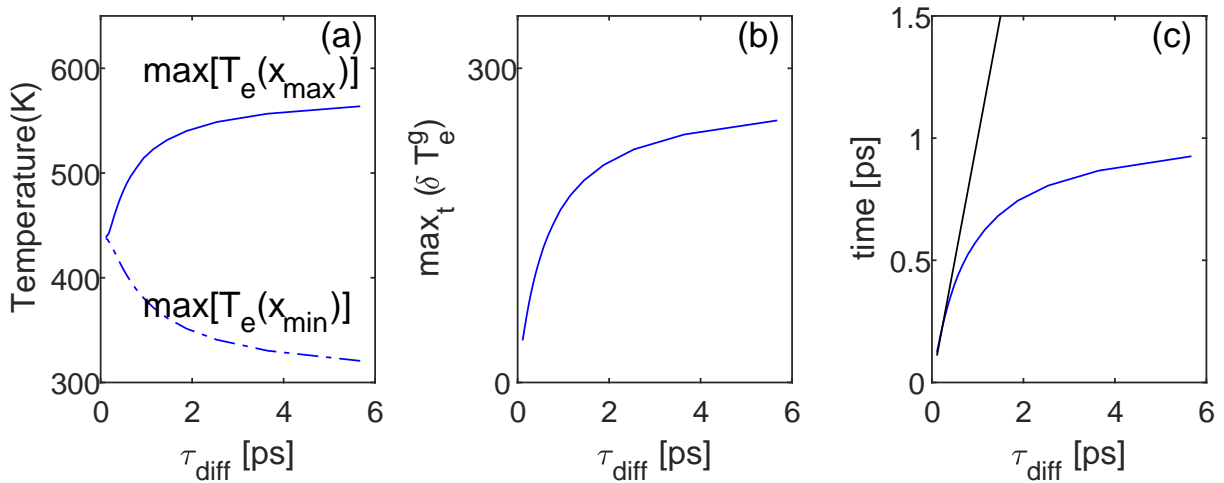}}
  \caption{(Color online) (a) Maximum values of $T_e(x_{max})$ (solid line) and $T_e(x_{min})$ (dash-dotted line) as function of diffusion time $\tau_{diff}$ for an infinitely thin metal film. (b) The corresponding maximal grating contrast. (c) $\tau_e^g$ (blue line) as a function of the diffusion time $\tau_{diff}$ compared with $\tau_{diff}$ itself (black line). }
  \label{fig:Te_max_min}
\end{figure}

In order to quantify the decay dynamics of the grating contrast, we plot in Fig.~\ref{fig:Te_max_min}(a) the temporal maximum of $T_e(x_{max})$ and $T_e(x_{min})$ for different values of $\tau_{diff}$. One can see that the dependence of the maximal values is strongest for short diffusion where $T_e(x_{max})$ drops and $T_e(x_{min})$ increases. Accordingly, the grating contrast $\delta T_e^g$ decreases for shorter diffusion times (Fig.~\ref{fig:Te_max_min}(b)). It however does not completely disappear, because of the delayed heating of the unilluminated regions - a transient temperature contrast builds up before the temperature at the non-illuminated regions.

We now define $\tau_g^e$ as the characteristic time scale in which the electron temperature contrast is erased. To determine $\tau_g^e$, we fit the decay stage of the grating contrast to an exponential function (see Fig.~\ref{fig:dynamicsTel}(g)-(i)). % \footnote{As shown below, the later stage of the dynamics does not exhibit an exponential decay. }. % , see details in Appendix~\ref{sec:numfit} - it avoids the latest part....
Potentially unexpectedly, one can see in Fig.~\ref{fig:Te_max_min}(c) that when the diffusion time is long, the grating is erased much faster than the diffusion time~(\ref{eq:taudiff}); this happens due to the spatially-local but temporally-nonlocal effect of $e-ph$ energy transfer which occurs roughly on a $1/\Gamma^{T_e}$ time scale. When the diffusion is faster, then, the grating erasure time approaches $\tau_{diff}$, as one might expect\footnote{Mathematically, it is indeed easy to see that in the extreme spatial nonlocal limit ($d \to 0$, hence, $\tau_{diff} \to 0$), the $e-ph$ coupling term is negligible with respect to the diffusion term, such that the dynamics of the electron temperature is dictated by the convolution between the temporally smeared source~(\ref{eq:U_NT_solution}) and the impulse response of the system (namely, the Green's function of the single temperature heat equation).}.

We now look again at the later stages of the dynamics shown in Fig.~\ref{fig:dynamicsTel}(g)-(i). Peculiarly, one can see that the rapid grating contrast decay slows down significantly. Additional numerical simulations of the complete eTTM (not shown) show that the diffusion proceeds at the phonon diffusion rate (as already shown in~\onlinecite{Maznev-Nelson-2011,ICFO_Sivan_metal_diffusion}); clearly, this occurs due to the coupling of the electrons to the phonons (the grating erasure indeed proceeds much more rapidly if the electrons are artificially decoupled from the phonons). Within the few picosecond range, the residual contrast is $\lesssim 1\%$, specifically, it is a few degrees for the local case but, as one might expect, it is much lower, a fraction of a degree, in the nonlocal case; naturally, the (residual) contrast decreases for a stronger electron thermal conductivity.

In order to complete the description of the heat dynamics, we also show the spatio-temporal dynamics of the phonon temperature. One can observe a behaviour reminiscent of that of the electron temperature in the local and nonlocal limits. Importantly, the actual phonon diffusion is faster than its intrinsic value because of the fast electron heat diffusion and the spatially-local $e-ph$ energy transfer. We also observe the non-zero residual phonon temperature grating contrast in the few picosecond regime; this residual phonon grating contrast is, however, much stronger than for the electron temperature - it is 85\%, 45\%, 20\% of the maximal phonon temperature rise, respectively, for the simulations in Fig.~\ref{fig:dynamicsTel}, specifically, at a few degree level.

\begin{figure}[H]
\centering{\includegraphics[width=18cm]{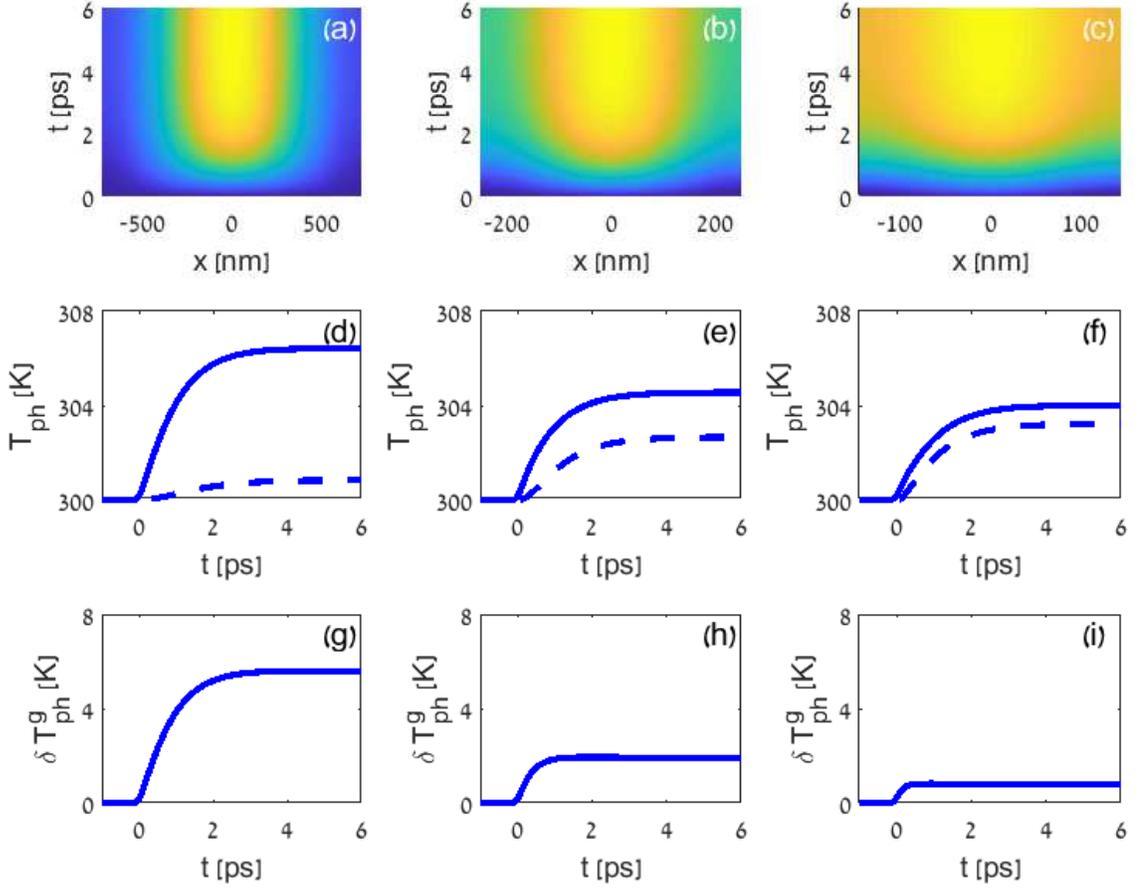}}
  \caption{(Color online) Same as Fig.~\ref{fig:dynamicsTel} for $T_{ph}$; temperature range plotted is $300$K to $308$K in all plots. } \label{fig:dynamicsTph}
\end{figure}

\section{Permittivity dynamics}

We use the temperature-dependent permittivity model described in~\onlinecite{ICFO_Sivan_metal_diffusion} to study the spatio-temporal dynamics of the metal permittivity $\epsilon_m(T_e,T_{ph})$ induced by the change in the temperatures. This dynamics thus determines the thermo-optic nonlinear response of the metal to the incoming light. Specifically, the probe pulse is centered at 850nm for which the contribution of interband transitions to the permittivity can be lumped into a single (temperature-independent) constant $\epsilon_\infty$. Accordingly, we refer below only to $\epsilon_m^{intra}(T_e,T_{ph})$.

\begin{figure}[H]
  \centering{\includegraphics[width=18cm]{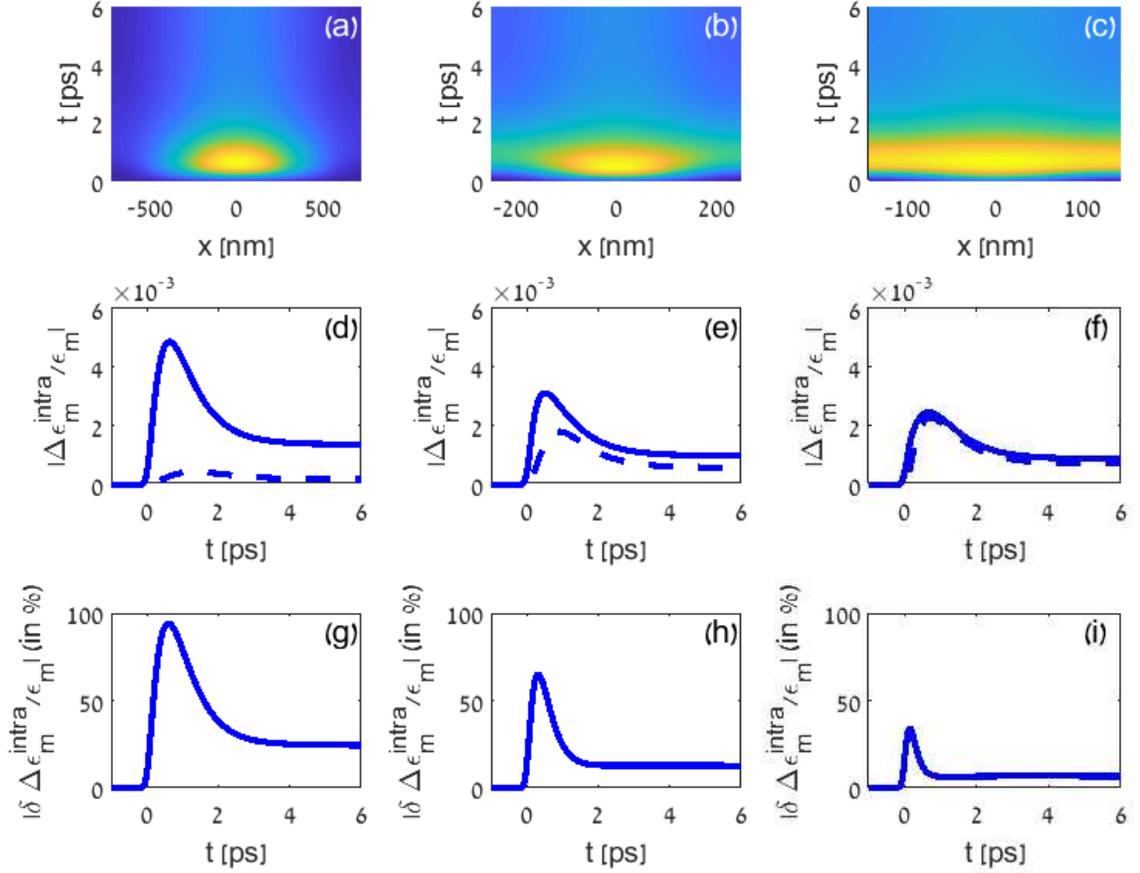}}
  \caption{(Color online) Same as Fig.~\ref{fig:dynamicsTel} for $\epsilon^{intra}_m(T_e,T_{ph})$; permittivity range plotted is $0$ to $6 \cdot 10^{-3}$ in all plots. } \label{fig:dynamics_epsilon}
\end{figure}

Fig.~\ref{fig:dynamics_epsilon} shows an initial rise of the permittivity due to the $T_e$ rise and a corresponding relative permittivity change below $1\%$ for the cases plotted in Fig.~\ref{fig:dynamicsTel} (compare to a $90\%$ relative change in the electron temperature). At later stages, the sensitivity to the phonon temperature becomes dominant such that for the $1\%$ change of $T_{ph}$, a $\sim 0.1\%$ change of the permittivity is obtained. This implies on much higher sensitivity of the permittivity to the phonon temperature, and a permittivity thermoderivative of $\sim 0.1/K$; such values are comparable to those found in the ellipsometry measurements described in~\onlinecite{PT_Shen_ellipsometry_gold}. % {\bf - 5\% FOR THE REAL PART, 2\% FOR THE IMAGINARY!. }
% As shown in~\cite{Gurwich-Sivan-CW-nlty-metal_NP} for the steady-state thermo-optic nonlinearity, the changes in the FIELD??! due to the real part are of second order... so maybe

It is interesting to note that the Au permittivity change occurs mostly due to the change in the imaginary part of the permittivity. This is similar to the theoretical and experimental observations in the CW case reported in~\onlinecite{SUSI,Gurwich-Sivan-CW-nlty-metal_NP,Un-Sivan-size-thermal-effect}. For Ag, the changes of the real part of the permittivity may play a non-negligible role~\cite{Gurwich-Sivan-CW-nlty-metal_NP,IWU-Sivan-CW-nlty-metal_NP}.

These well-understood aspects of the problem have important implications to the spatio-temporal dynamics of the metal permittivity. First, one can see that heat diffusion reduces effectively the maximal change of the permittivity - this is a result of the non-locality mediated heating at the illuminated regions. In contrast, the permittivity change at the few picosecond time scale is nearly unaffected by diffusion - this is because the permittivity change is dominated by the phonon temperature (see Fig.~\ref{fig:dynamicsTph}(d)-(f)). Second, while it seems from Fig.~\ref{fig:dynamics_epsilon}(c) and~(f) that the permittivity grating contrast (defined as $\delta \Delta \epsilon_m = \epsilon_m(t,x_{max}) - \epsilon_m(t,x_{min})$) vanishes in the spatial non-local limit, a more careful look (Fig.~\ref{fig:dynamics_epsilon}(i)) shows that the permittivity grating contrast is reduced, but still significant (a few tens of percent) - this is due to the finite time required for heat to diffuse to the unilluminated regions. In that sense, the rise and decay time scales of the grating contrast are faster than those of the maximal permittivity change itself, a fact that could be exploited for ultrafast switching applications~\footnote{A same behaviour is observed for the temperature dynamics (Fig.~\ref{fig:dynamicsTel}); however, due to the log scale used in Fig.~\ref{fig:dynamicsTel}(g)-(i), this behaviour is somewhat obscured. }. Nevertheless, the maximal permittivity contrast is $\sim 35\%$ of the maximal permittivity change in the non-local case (Fig.~\ref{fig:dynamics_epsilon}(i)), whereas it is only $\sim 6\%$ at the few picosecond regime. This level is far higher than the relative residual electron temperature grating ($\sim 0.1\%$, compare to Fig.~\ref{fig:dynamicsTel}(f) and (i)). Furthermore, although the residual permittivity contrast is small, it persists for much longer than the short transient peak. As a result, the reflectivity from the permittivity grating will persist beyond the initial picosecond, thus, putting the applicability of TBGs in metal films for ultrafast switching configurations based on self-erasure (see e.g., in~\onlinecite{Sivan-COPS-switching-TBG}) in question.

Overall, the above results show that the permittivity change depends on the interference patten (i.e., on the period) even though the local field hence absorbed power density are the same for all cases. More generally, the nonlinearity at short times (and to a lesser extent, also in the picosecond time scale) changes with the spatial frequency contents of the illumination (by more than a factor 2). This implies that in addition to the temporal nonlocal nature of the metal response (see~\onlinecite{Biancalana_NJP_2012}), the nonlinear thermo-optic response of metals has also a {\em spatial nonlocal} aspect, an aspect which may contribute to the multiplicity of values assigned for the nonlinearity across the literature (see, e.g.,~\onlinecite{Boyd-metal-nlty,NL_plasmonics_review_Zayats_2012}).

\section{Discussion}\label{sec:discussion}
Our simulations show that the heat diffusion in metals determines the grating erasure time for $\tau_{diff} \ll 1/\Gamma^{T_e}$ ($\sim 1.25 ps$ in Au). In particular, diffusion determines directly the build up time of the heat in the minimum illumination points and mediates the heating level of the strongly illuminated regimes. This enables metal to overcome conventional limitations on the nonlinear response~\cite{Zheludev-metal-nlty} - the response is both fast and strong. When the diffusion time is longer, the grating erasure time is dominated by spatially-local but temporally-nonlocal effects ($\Gamma^{T_e}$).

Our analysis, and especially the identification of the nonlocal nature of the permittivity change, is also essential for understanding the nonlinear thermo-optic response of metals. In particular, it shows that the nonlinear thermo-optic response depends not only on the pump and probe wavelengths, the illumination intensity and duration (see e.g.~\onlinecite{Boyd-metal-nlty}), but also on the spatial distribution of the pump, namely, its spatial (in)homogeneity.

From the applied physics perspective, our results show that most of the grating contrast disappears on a sub-picosecond time scale. However, even then, there is a residual grating contrast that persists for many tens of picosecond even for very strong heat diffusion (nonlocal case), i.e., for very short-scale thermal inhomogeneities (the induced grating period in the current case). Therefore, TBG applications based on metals might be more constrained compared to transient dynamics of free carriers in semiconductors~\cite{Sivan-COPS-switching-TBG}.

The results shown in this manuscript are robust to the various approximations made in our model. Specifically, the phonon diffusion does not change the dynamics described above (indeed, it is a much weaker effect compared with the electron-induced phonon diffusion); there is also only weak sensitivity to the pump duration and to the metal layer thickness. Indeed, previous studies (e.g.,~\onlinecite{Maznev-Nelson-2011,Lalanne_Narang_Nat_Comm_2017,ICFO_Sivan_metal_diffusion} showed that accounting for the finite thickness gives rise only to modest quantitative changes in the temperature dynamics. Yet, our result might be sensitive to anisotropy in the diffusion and to structuring~\cite{Zayats_Nonlocal_nlty,Lalanne_Narang_Nat_Comm_2017}.

Finally, as the inclusion of heat diffusion in our model was done phenomenologically, in a manner compatible with a classical description of the underling physics, it is clear that a self-consistent model has to be derived from a more fundamental point of view, even going beyond the Boltzmann equation to a Density-Matrix formulation. Such a model might reveal some subtleties that the current model did not capture, and may even provide first theoretical predictions for ballistic (i.e., faster-than-diffusion) transport. These effects were envisioned theoretically, but so far studied with phenomenological models that do not account for the non-thermal energy dynamics (Eq.~(\ref{eq:dynamicsNT}); see e.g.,~\onlinecite{Hohlfeld_Matthias_2000,Ballistic_e_diffusion,Ivanov_Phys_Rev_Appl_2015}). % - yang colorado.. beraun,
Experimentally, there have been some studies of vertical heat transport based on a configuration in which the pump and probe are incident on opposing interfaces of a metal film (see~\onlinecite{Brorson1987,Hohlfeld_Matthias_2000,Planken_buried_gratings_2018}), however, for lateral heat transport, there has been no evidence for ballistic diffusion even when using femtosecond temporal resolution, see~\onlinecite{ICFO_Sivan_metal_diffusion}.

% Brorson - 50,100,200,300 nm thick films - fits with a linear curve and gets an offset of 2 delta_{skin}; could have been fit with a parabolic relation (diffusion), especially for the low thicknes. Matthias provided more conclusive evidence os the latter picture (looking at 10,20, ...,100nm thicknesses ans showing a qualitatively different behaviour for the thicker layers. Plancken did the same with better time resolution.

However, thinner structures, combined with advanced ultrafast techniques applied to plasmonic systems (see, e.g., ), might allow one to observe ballistic transport.

Convincing models and measurements of heat transport in this regime will also be useful for resolving arguments regarding charge transfer on the femtosecond scale which are relevant for non-thermal (the so-called ``hot'') electron studies, and their application for photocatalysis and photodetection~\cite{Dubi-Sivan,Dubi-Sivan-Faraday,dyn_hot_e_faraday_discuss_2019} and should eventually be combined with models of additional transport effects associated with interband transitions (see e.g.,~\onlinecite{Shalaev_LID1,Shalaev_LID2}).

Finally, our findings motivate the study of heat and charge diffusion in other systems, e.g., for materials with nearly-instantaneous thermalization such as graphite or graphene~\cite{Graphene-switching-Hendry,Petek_PRX_2017} and other 2D materials~\cite{Klaas_hBN,Klaas_graphene} on one hand, as well as for semiconductors such as GaAs for which the thermalization is slower~\cite{Petek_Si_1998,Euser_Thesis,Euser_Vos_JAP_2005}.

% Our findings exemplify that there are quite a few aspects of ultrafast heat diffusion which deserve further research. Another poorly-explored domain for heat diffusion is the several picosecond time-scales where significant acoustic dynamics occurs~\cite{Pelton_acoustics,Haim_acoustics,Crut_acoustics}.

% {\bf refer to Haim's studies with high t resolution.}

% \cite{Ballistic_e_diffusion} tries to say something about nonthermal electrons (ballistic movement) without modelling them... anyhow, this paper seems to confuse the concepts and/or double count them.

\bigskip

{\bf Acknowledgements.} The authors thank A. Block, P. Y. Chen, S. Sarkar and I. W. Un for many useful discussions. We are grateful to an anonymous reviewer that pointed out a flaw in a previous version of this work~\cite{Sivan_Spector_metal_diffusion_wrong}. Y. S. was partially supported by Israel Science Foundation (ISF) Grant No. 899/16.

% Maznev - relatively unknown result

\bibliographystyle{unsrt}
% PC -
% \bibliography{C:/Users/sivanyon/Documents/Research/my_bib}
% laptop
% \bibliography{C:/Users/Yonatan/Documents/Research/my_bib}

\providecommand{\latin}[1]{#1}
\providecommand*\mcitethebibliography{\thebibliography}
\csname @ifundefined\endcsname{endmcitethebibliography}
  {\let\endmcitethebibliography\endthebibliography}{}

\end{document}